# Toward Effective Multi-Domain Rumor Detection in Social Networks Using Domain-Gated Mixture-of-Experts

*Mohadeseh Sheikhqoraei[a], Zainabolhoda Heshmati[b], Zeinab Rajabi[c], Leila Rabiei[d]*

[a] School of Intelligent Systems, University of Tehran, Iran, m.qoraeii@gmail.com

[b] School of Intelligent Systems, University of Tehran, Iran, zheshmati@ut.ac.ir

[c] Zeinab Rajabi, Hazrat-e Masoumeh University, Qom, Iran, z.rajabi@hmu.ac.ir

[d] Leila Rabiei, Iran Telecommunication Research Center (ITRC), Tehran, Iran, l.rabiei@itrc.ac.ir

## Abstract

Social media platforms have become key channels for spreading and tracking rumors due to their widespread accessibility and ease of information sharing. Rumors can continuously emerge across diverse domains and topics, often with the intent to mislead society for personal or commercial gain. Therefore, developing methods that can accurately detect rumors at early stages is crucial to mitigating their negative impact. While existing approaches often specialize in single-domain detection, their performance degrades when applied to new domains due to shifts in data distribution, such as lexical patterns and propagation dynamics. To bridge this gap, this study introduces PerFact, a large-scale multi-domain rumor dataset, comprising 8,034 annotated posts from the X platform annotated into two primary categories: rumor (including true, false, and unverified rumors) and non-rumor. Annotator agreement, measured via Fleiss' Kappa ($\kappa$ = 0.74), ensures high-quality labels.

This research further proposes an effective multi-domain rumor detection model that employs a domain gate to dynamically aggregate multiple feature representations extracted through the Mixture-of-Experts method. Each expert combines CNN and BiLSTM networks to capture local syntactic features and long-range contextual dependencies. By leveraging both textual content and publisher information, the proposed model classifies posts into rumor and non-rumor categories with high accuracy. Evaluations demonstrate state-of-the-art performance, achieving an F1-score of 79.86% and an accuracy of 79.98% in multi-domain settings.

## 1- Introduction

Social networks, with their large user bases and complex interactions, have become an integral part of modern social life. These platforms generate vast amounts of textual data that can be leveraged for various analytical purposes. In particular, sentiment analysis[1],[2], rumor detection[3], and social resilience assessment[4] within social networks serve as crucial tools that provide researchers and decision-makers with deep insights into user behavior and social dynamics. By extracting behavioral and content patterns, these analyses enable the rapid and accurate identification of social phenomena, thereby enhancing information management processes and informing effective social policy-making. Social media are considered important platforms for the spreading of news and information. Due to the ease of sharing content, rumors also spread easily through these networks. Sometimes, rumors are just misleading information shared among people, while at other times, they are intentionally propagated and can cause irreparable damage to organizations, institutions, individuals, governments, and even the general public. They can increase social anxiety, reduce productivity and efficiency, paralyze the economic cycle, and foster distrust, pessimism, and suspicion in society[5]. Previous studies have used different definitions to describe rumors. For instance, in some studies, rumors have been defined as information perceived to be false[6], whereas most of the literature defines rumors as "circulating unverified informational statements"[7]. In this study, we refer to rumors as content that remains unverified at the time of dissemination.

In recent years, numerous studies have been conducted worldwide on rumor detection using artificial intelligence methods. Events such as the 2016 U.S. elections and the COVID-19 pandemic have increased the attention of researchers to this field. Moreover, the emergence of advanced and novel deep learning models, with their superior performance and high accuracy, has led to their use in achieving better results across various applications, including rumor detection. These rumors are disseminated across diverse domains such as politics, economics, health, and more. Significant differences often exist between rumors from different domains, such as the vocabulary used within each domain, which results in poor performance of those methods when applied to other domains, namely domain shift[8]. For example, commonly used words in the entertainment domain include "actor" and "cinema", while in military news, "army" and "suppression" are frequently used words. Most existing methods focus on single-domain rumor detection and experience a noticeable drop in accuracy when encountering rumors from a new domain. Retraining the model on new data is not ideal in practice, as there will always be new data that the model has not previously encountered. Additionally, retraining the model from scratch with new datasets is not practical, as data collection and retraining are time-consuming and costly, undermining the primary goal of rapid rumor detection. Therefore, in this study, we aim to propose a multi-domain rumor detection approach to enable quick detection of rumors, mitigate their harmful effects, and reduce the costs associated with model retraining and data collection.

It is important to explain the term "rumor detection" as it is often confused with another closely related term, "rumor verification." The goal of rumor detection is to determine, based on relevant information shared by users, whether a post is classified as a "rumor" or "non-rumor." In contrast, "rumor verification" involves determining the truthfulness of a suspected rumor (true, false, or unverified)[9]. In This paper we focus on rumor detection. According to the defined concept of a rumor, the problem addressed in this research is the development of a multi-domain Persian rumor detection method based on deep learning approaches, as well as the creation of a multi-domain Persian dataset on the X platform. Using this dataset, the proposed method aims to mitigate the harm caused by the widespread dissemination of rumors. Specifically, in this study, the collected dataset undergoes preprocessing and embedding before being input into a deep classification model that employs the Mixture-of-Experts approach. After feature extraction and deep model training, posts are categorized as either being rumors or non-rumors.

Deep learning methods are good at extracting relationships between words and representing complex features, allowing for more accurate classification of intricate patterns. However, these methods require large datasets to achieve desirable accuracy and avoid overfitting on training data. In recent years, several studies have been conducted on English datasets, such as PHEME[10]. However, a significant challenge in detecting rumors and misinformation in low-resource languages is the lack of comprehensive and suitable datasets for training. Compared to English, research in these languages has seen limited progress. While some studies in Persian have collected datasets[11], [12], [13], these datasets are insufficient in terms of size and comprehensiveness for detecting multi-domain rumors on Persian social media using state-of-the-art deep learning methods.
Rumors are continuously spread across various topics, and training a model solely on data from a specific domain or event diminishes its effectiveness in predicting rumors in other domains or future events. Based on the conducted reviews, it seems that no prior research has focused on multi-domain rumor detection in Persian. Consequently, the contributions of this research are as follows:

- **Creation of a Comprehensive Rumor Dataset**: We curated a diverse Persian rumor dataset comprising 3,839 rumor posts and 4,198 non-rumor posts spanning seven distinct domains, all sourced from X. This dataset addresses a critical gap in Persian-language resources for rumor detection.

- **Introduction of a Novel Hybrid Multi-Domain Model:** We proposed a new hybrid model tailored for multi-domain Persian rumor detection. Our approach combines a Mixture-of-Experts architecture with deep learning components, including CNN and LSTM, while integrating metadata to significantly boost detection accuracy and robustness.

- **Advancement in Multi-Domain Persian Rumor Detection:** To the best of our knowledge, this work is the first to comprehensively tackle the challenge of multi-domain rumor detection in Persian, establishing a foundational benchmark for future research in this area.

This study aims to address the following research questions:

- **RQ1:** How effectively can a multi-domain deep learning model detect rumors across diverse domains, particularly within a low-resource language context?
- **RQ2:** How accurate is the collected X platform dataset when applied to the proposed research model?

To answer these questions, the study proposes a deep learning-based model and a multi-domain dataset for Persian rumor detection.

The remainder of this paper is organized as follows: Section 2 reviews the related literature. Section 3 introduces and describes the dataset collected for this study. Section 4 outlines and explains the proposed method in detail. Section 5 presents the results of the proposed method and compares them with findings from other relevant studies. Section 6 discusses the study's limitations and outlines directions for future research. Finally, Section 7 summarizes the conclusions and contributions of this study.

# 2- Related Works

In this section, we briefly review its relevant work from five perspectives: (1) Advancements in Rumor Detection Methods, (2) Feature Extraction Paradigms, (3) Domain Adaptation Strategies (4) Persian Rumor Detection.

## 2-1- Advancements in Rumor Detection Methods

Many methods have been proposed to address the challenges of detecting rumors and false information. Earlier studies primarily relied on machine learning approaches that were dependent on feature engineering for identifying rumors[14], [15]. However, with the increasing availability of data and the superior performance of deep learning methods, these advanced techniques have become the primary choice today [16], [17], [18], [19], [20]. Wu et al.[14] proposed using a graph kernel-based SVM classifier for rumor detection. They highlighted that the propagation patterns of true and false rumors differ, making the random walk graph kernel particularly effective for identifying false rumors. Della Vedova et al.[15] introduced a hybrid technique for fake news detection. Their approach combined social context-based methods, including logistic regression and harmonic Boolean labeling for crowdsourced feedback (used when an item has one or more responses), with content-based methods (used when no feedback is available). They implemented their method in a chatbot operating on Facebook Messenger, which accepts Facebook post URLs from users and determines whether the posts are true or fake.

Yang et al.[16] proposed a novel approach called TI-CNN, which integrates textual and image information along with their corresponding explicit and implicit features to detect fake news. Chauhan et al.[17] developed a deep learning framework incorporating neural networks and LSTM, focusing on sequential data classification for prediction. They used the GloVe technique to represent each word as a vector for word embeddings. Lin et al.[18] proposed a graph-based autoencoder method for rumor detection, consisting of three components: an encoder, a decoder, and a detector. They used the GCN kernel module to update node representations and encode the graph structure. The model was trained with a joint learning strategy involving the encoder, decoder, and detector to capture textual, propagation, and structural information simultaneously. Liu et al.[19] used GNNs to learn user correlation representations from a bipartite graph, describing correlations between users and source posts, and propagation representations with a tree-structured format. They combined these learned representations for rumor classification. Given that malicious users aim to disrupt detection models, they developed a greedy attack scheme to analyze the cost of three adversarial attacks: graph attack, comment attack, and joint attack. Since transformers can capture dependencies in long sequences, Khoo et al.[20] employed the multi-head attention mechanism in transformers to model the relationships between posts. They differentiated the community's responses to real and fake claims on microblogs for rumor detection. The study notes that existing advanced methods are often tree-based, modeling conversation trees. However, on social media, a user responding to a post may address the entire topic rather than a specific user. Therefore, the interdependence of user comments across different discussion threads should also be considered. The multi-head attention mechanism in transformers was used to extract features from main posts and retweet comments, calculate correlations among them, and assign weights to these correlations, allowing user comment features to serve as a foundation for more precise rumor detection. Beseiso & Al-Zahrani[21] created a fake news detection model that mixes bidirectional LSTM with 2D CNNs to catch patterns in word groups (2-grams, 3-grams, 4-grams). Their ensemble model hit 97.3% accuracy, beating earlier model, especially when using longer word sequences, pre-trained embeddings, and varied tokenization.

## 2-2- Feature Extraction Paradigms

Zhang et al.[22] distinguish between user-based, content-based, and context-based approaches for feature extraction. User-based features refer to the characteristics of social media users who create or share news, such as the number of posts, membership date, number of friends/followers, etc. Other related information pertains to the level of activity and presence of individuals within social circles. For example, they can be calculated from the number of posts as well as the number of followers and followings [23],[24].

Content-based features are typically extracted directly from the text. Most studies have utilized such features. By performing in-depth analysis of news content, we can analyze linguistic patterns and writing styles for both real and fake content, then use the most distinctive features for rumor detection[22]. Bahuleyan et al.[25] proposed a set of words made up of signal words to determine people's attitudes toward false information, categorizing them into meaningful groups such as "belief," "denial," "doubt," "report," "knowledge," and "refutation." These words serve as reliable indicators of the writer's stance because the presence of belief or knowledge-related words may indicate the writer’s agreement with the news. Syntactic features extracted from sentences often identify existing patterns from grammatical structures. For instance, Zeng et al.[26] used part-of-speech tagging (POS), while Ghanem [27] used named entity recognition (NER). Prakash et al. [28], used TF-IDF to represent the frequency of words. They combined TF-IDF with the RoBERTa pre-trained model, improving classification accuracy, especially in the denial class, which suffered from data scarcity. Palani et al.[29], proposed a model for detecting fake news in its early stages using BERT, analyzing both textual and visual content of news articles. The proposed CB-Fake model collects textual and visual features to learn an advanced multi-dimensional feature representation. This model uses CapsNet to extract further visual features from the image. It also employs BERT to depict context-based textual features from news articles. These features are then combined and sent to a classification layer to determine whether the news is fake or real. Ghanem et al.[30], examined several style-based features, such as the frequency of question and exclamation marks, sentence length, and the ratio of capital letters in the text, reporting significant advancements. Social media-specific features are another content-related indicator for rumor verification, including mentions, emojis, and hashtags[31].

Context-based features are extracted by considering the related information around fake news or posts in social media. Contextual features include the structure of information dissemination on social media and the reactions of other users to the news. Network-based features are used to model information related to the network in which news is shared. For example, the dissemination structures, spreading patterns, and subgraph features where news is shared, such as density and clustering coefficient, are considered[22]. According to [32], most studies that implement network-based features focus on using statistics in dissemination patterns, such as the number of reposts and publication time. Other studies have focused on modeling temporal features of news dissemination [23], [33]. For example, Kwon et al.[23] created several networks based on friendship status between users and dissemination patterns, extracting features based on the clustering coefficient and degree of such networks. Castillo et al.[34] used a combination of content-based and context-based features such as sentiment score, text length, and changes in certain user features throughout the message lifecycle to evaluate the credibility of posts. They stated that trending topics tend to include URLs and deep dissemination trees. Furthermore, reliable news is usually published by writers who have previously written numerous messages and have many reposts.

### **2**-3- Domain Adaptation Strategies

Single-domain fake news detection focuses on a specific domain, such as politics or health, but often faces challenges due to the limited availability of data in that domain. In contrast, cross-domain detection specifically utilizes data from other domains to improve performance in a target domain. Multi-domain approaches leverage data from multiple domains to enhance overall performance, though they may also encounter the seesaw phenomenon, where improvements in one domain come at the cost of reduced performance in others[35].

Nan et al.[36] created a dataset called Weibo21, containing real news from nine different domains. Their proposed multi-domain model, MDFEND, detects fake news using an ensemble method.

Wang et al.[37] introduced a model called Event Adversarial Neural Network (EANN) to detect multi-domain fake news in emerging events. This model can extract invariant features of events, enabling fake news detection for new events. In this model, a multi-modal feature extractor is responsible for extracting textual and visual features from posts. This component collaborates with the fake news detector to learn representations for fake news detection. The event discriminator removes event-specific features and retains shared features across events.

Chen et al.[38] suggest that news from a source in one domain may semantically interact with other domains. Therefore, detecting fake news using only one domain might miss broader meanings and affiliations (e.g., connections to other domains). To address this, they proposed a new model called FuzzyNet, which overcomes these limitations using a fuzzy mechanism.

Huang et al.[39] proposed a multi-modal fake news detection model, MMCFND, to address the limitations of text-based methods. They used image-text alignment learning and adversarial learning to align the two modalities in a shared semantic space. An ensemble system was also integrated to enhance generalization in multi-domain fake news detection.

Li et al.[40] addressed the challenge of fake news detection across multiple domains, where traditional methods struggle due to domain shift. Their approach combines multiple BERT models, multilayer CNNs, and attention mechanisms for richer feature extraction in both short and long texts, along with a domain localization module using sentiment analysis for precise targeting. They also designed a fast model for timely detection, which together significantly improves detection performance across multiple domains

Ghayoomi and Mousavian[41] compared the results of their proposed model in four different scenarios. They noted that training a model with a dataset in a language similar to the test data, but not from the same domain, has a slightly positive impact on model performance. This is because the model learns linguistic features of the target language but may underperform due to a lack of domain-specific knowledge.

## 2-4- Persian Rumor Detection

In recent years, technology in Iran has accelerated information dissemination. Due to the prevalence of the rumors published in Persian, several studies have focused on their detection[11], [41],[42]. Zamani et al.[11] created a Persian dataset with 783 rumor posts from two websites, adding an equal number of non-rumor posts. Using features like number of links, likes, join date, and followers, they tested methods like Decision Tree and Naïve Bayes to identify common words in Persian rumors. Jahanbakhsh-Nagadeh et al.[42] proposed a content-based model for early detection of Persian rumors on X and Telegram. The model combines semantic, pragmatic (e.g., speech acts), and syntactic information. Text embeddings were created using ParsBERT and CapsNets, enriched with extracted pragmatic and syntactic features for improved verification. Ghayoomi and Mousavian[41] used XLM-RoBERTa and a parallel CNN classifier to detect fake Persian news about COVID-19. They used three different datasets to evaluate the proposed model: an English dataset, a Persian fake news dataset (general domain), and a Persian dataset for COVID-19 fake news detection. They stated that merely using the English COVID-19 fake news dataset in cross-lingual transfer learning does not improve results, as the model does not learn about the linguistic features of the Persian language.

## 3- Dataset

Supervised deep learning methods require large labeled datasets. While efforts have been made to collect Persian rumor or fake news datasets[11],[13], there remains a significant need for a well-labeled multi-domain rumor dataset in Persian and expanding the size of available datasets could greatly enhance the performance of AI-based methods. For this purpose, we present "PerFact", a fact-checked Persian rumor dataset containing 3,839 rumor posts and 4,198 non-rumor posts.

### 3-1- Data collection

In the process of collecting relevant datasets, a collection of credible documents was first created using reputable official news agencies such as ISNA[1], IRNA[2], and others, as well as fact-checking websites like Factnameh[3] and Factyar[4]. This collection includes news documents related to ten different events in various domains, including politics, society, science and technology, health, entertainment, military, and economics. These events have been discussed and shared on X. After studying the related documents, keywords associated with each event were extracted. For example, in the case of the rumor about lithium extraction from Lake Urmia, keywords like "lake," "earthquake," "Khoy," "explosion," "HAARP," "extraction," and "Urmia" were among the relevant terms used for searching posts. Subsequently, the relevant posts for these events were collected using X's application programming interface (API), based on the extracted keywords. These posts were posted between January 2020 and August 2024. In addition to the tweet text, other information, such as metadata, user information of the tweet author, and comments from other users, was also collected.

[1] www.isna.ir

[2] www.irna.ir

[3] www.factnameh.com

[4] www.factyar.com

### 3-2- Data and Domain annotation

Labeling can be considered part of the data preprocessing stage, where assigning the relevant label to each sample allows the model to make more accurate predictions. Using fact-checking websites and news agencies, a collection of documents was gathered to identify events. Annotators, with the help of these documents that extensively describe the events and verify their authenticity, determined the relevant label for each data sample, categorizing them into one of four classes. The number of labeled data points in each category is as follows: 4,198 posts in the non-rumor class, 1,725 posts in the true rumor class, 1,656 posts in the false rumor class, and 458 posts in the unverified rumor class.

The following describes the labeling process and provides examples from the dataset.

a) Non-rumor posts

These posts are contextually related to the rumor topic but do not convey any true or false information about it. Examples include:

Providing additional information instead of addressing the specific rumor. For example, instead of discussing the rumor about “Assassination of Judge Salavati”, the tweet might talk about his personality traits or lifestyle.

Sharing personal opinions, emotions, or memories related to the topic, such as sympathizing with victims of an incident, offering condolences, calling for action against perpetrators, criticizing individuals, or quoting religious verses or famous sayings relevant to the event.

Asking other users for opinions about an event.

b) True rumor posts

This category contains information that is entirely accurate and confirmed by news agencies or fact-checking websites.

c) False rumor posts

These posts convey statements or statistics that are incorrect or refuted by credible evidence.

d) Unverified rumor posts

This category includes news and rumors for which reliable evidence neither confirms nor refutes their accuracy. The available information is insufficient to classify them as true or false. Typically, these posts involve quoting, describing hearsay or observations without proof, or speculating about the intentions and objectives of organizations or other entities.

*Table 1- Samples of different labels*

| | |
|---|---|
| **Non-rumor** | |
| Look, there's this damn principle: the burden of proof is on the claimant. So, the one who says someone's been killed has to bring the evidence, not a bunch of people saying the other one should prove that Salavati isn't dead! | ببینید، یه اصل کوفتی وجود داره که: بار اثبات ادعا با مدعیه در نتیجه، اونی که میگه فلانی کشته شده باید مدرک ارائه بده، نه اینکه یه مشت بگن فلانی مدرک بیاره که صلواتی نمرده! |
| **True** | |
| There was a rumor that Salavati was assassinated. We said it was false. And it was false. | شایعه کردند صلواتی ترور شده، گفتیم دروغه. دروغ بود. |
| **False** | |
| There's a heavy security presence at Judge Salavati's residence; it seems the news of his assassination by partisans was true and confirmed! | جو امنیتی شدیدی در محل سکونت قاضی صلواتی هست؛ گویا خبر کشته شدنش توسط پارتیزان‌ها درست بوده و خبر موثق است! |
| **Unverified** | |
| Salavati's house is definitely not in Niavaran; it's on Zafar Street, near Shariati, in Sabr Alley | اصلاً صلواتی خونه‌اش نیاوران نیست، ظفر می‌شینه نزدیک شریعتی کوچه‌ی صبر |

As noted in table 1, this research aims to detect rumors by classifying them into two categories: rumor and non-rumor. Most previous efficient methods, which will serve as baselines for evaluating the proposed model, also treat rumor detection as a binary classification problem. Therefore, posts in the true, false, and unverified categories are combined as the rumor class in the proposed model.

### 3-3- Data Analysis

Figure 1 displays word clouds for the top words in different domains, illustrating the distinct distribution and frequent terms across these domains. This highlights the linguistic variations specific to each domain and the importance of building a multi-domain rumor dataset.

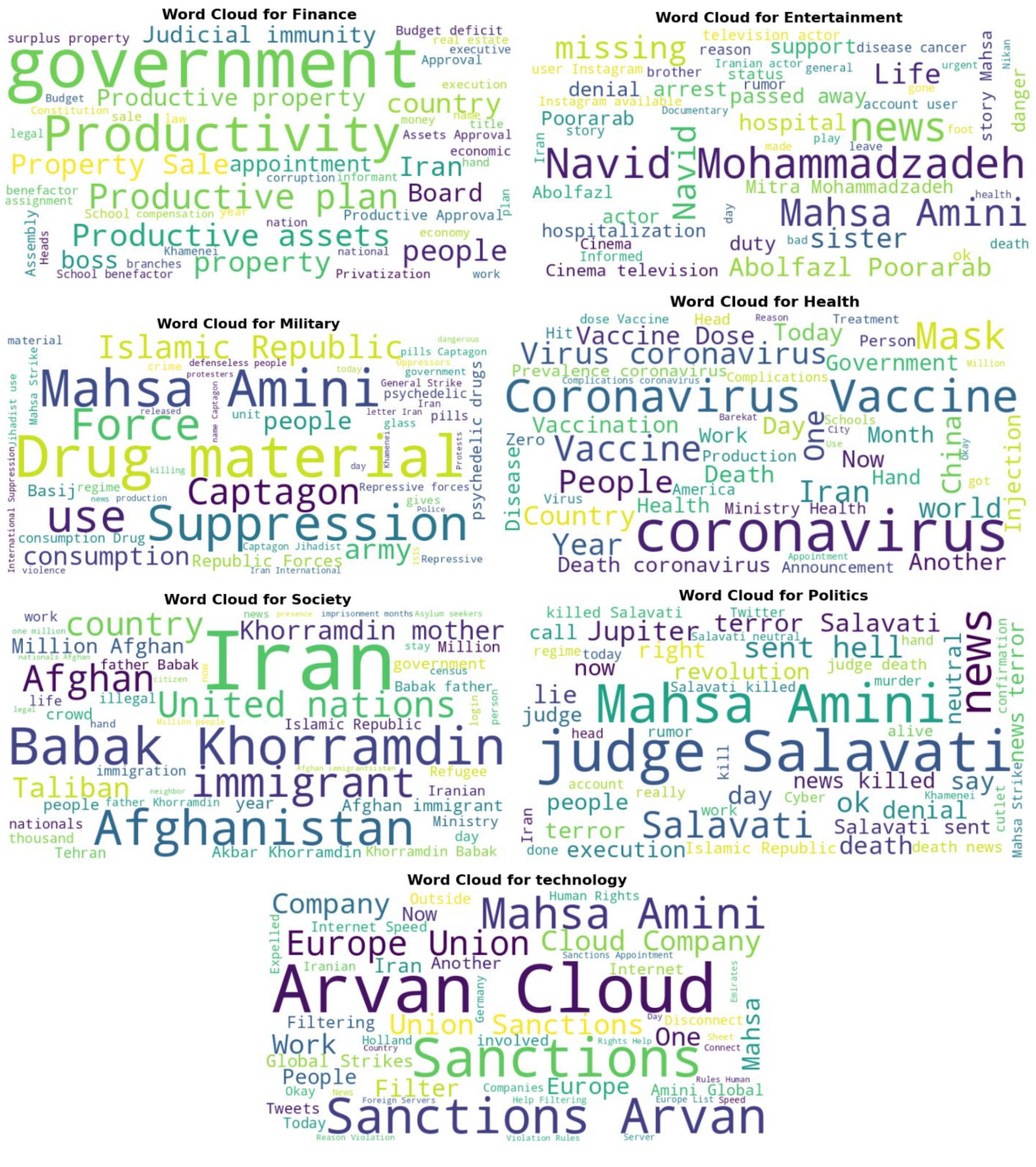

*Figure 1 - Word Clouds of the seven Domains. The Technology domain is characterized by the high frequency of the word Arvan Cloud, whereas the Health domain is dominated by the word coronavirus, reflecting the domain-specific word distributions within the corpus.*

The collected dataset requires domain labels as well. Following [36] and reviewing the classifications used by news websites like ISNA, events were categorized into nine general domains: politics, science and technology, economy, society, health, entertainment, military, education, and disasters. With the help of a three-member team, each event was assigned to one of these nine domains.

Finally, before preprocessing, the dataset contained 8,037 posts across seven domains: politics, science and technology, economy, society, health, entertainment, and military, with labels grouped into two categories: rumor and non-rumor for use in the MDRD model.

*Table 2-Number of samples per domain and category*

| **Domain** | Politics | Technology | Economy | Society | Health | Entertainment | Military | Total |
|---|---|---|---|---|---|---|---|---|
| **Rumor** | 1,280 | 210 | 409 | 580 | 1,133 | 74 | 153 | 3,839 |
| **Non-rumor** | 1,275 | 157 | 853 | 820 | 1,060 | 19 | 14 | 4,198 |
| **Total** | 2,555 | 367 | 1,262 | 1,400 | 2,193 | 93 | 167 | 8,037 |

### **3**-4- Human verification

To improve accuracy and assess impartiality in this research, the labeling process for the event "Assassination of Judge Salavati" with 2,449 samples from the dataset was conducted with the collaboration of a three-member team. To confirm and validate the labels assigned by the annotators, a measure is needed to calculate the level of agreement among them. One of the well-known metrics for measuring agreement among annotators is Fleiss' Kappa. This metric is calculated for group annotation and was used in this study. The five categories considered for Fleiss' Kappa calculation are: removal, non-rumor, true rumor, false rumor, and unverified rumor. The level of agreement between annotators was found to be 0.74, using the Python Statsmodels library to calculate this metric. The interpretation of various Fleiss' Kappa values is shown in the table 3[43]:

*Table 3- Interpretation of Fleiss' Kappa Values*

| Fleiss' Kappa Value | Annotator Agreement Level |
|---|---|
| 0.00 > | Weak agreement |
| 0.00 – 0.20 | Partial agreement |
| 0.21 – 0.40 | Fair agreement |
| 0.41 – 0.60 | Moderate agreement |
| 0.61 – 0.80 | Substantial agreement |
| 0.81 – 1.00 | Almost perfect agreement |

According to Table 3 and its interpretation, the annotators in this study fall within the "substantial agreement" range, indicating a high level of agreement.

## 3-5- Data preprocessing

The collected data often have an inappropriate structure, extra information, and an unsuitable size for processing. Data preprocessing includes a set of tasks designed to clean and standardize textual data. The goal of these tasks is to reduce noise and improve data quality, allowing models to learn more effectively and produce more accurate results. To appropriately use the dataset, preprocessing is carried out along with the following Steps[44]:

- **Unicode Correction**
  Unicode is a standard encoding system that uniformly represents text from various languages. Unicode correction involves addressing issues such as incorrect character encoding, managing special characters, and normalizing different representations of a character.
- **Normalization of Common Emojis**
  frequently used emojis in the dataset, with more than 10 occurrences across the entire dataset, were replaced with their equivalent descriptions. For example, the emoji 😆 was replaced with the word "laugh."
- **Data Cleaning**
  The "Clean-Text" library is used where the quality and consistency of textual data are important for model building. Using the "Clean-Text" library, the following actions were taken for data cleaning:

- Removal of extra space, line breaks, URLs and email addresses, phone numbers
- Removal of other unnecessary emojis and special characters

Additionally, the Regex library was used to remove symbols like "#" before hashtags.

- **Data Normalization**
  To normalize the Persian textual data, the "Hazm" library was used. The primary goal in this step is to unify and standardize characters to have more control over the results of advanced text processing. The normalization process is carried out using the normalize function of the Normalizer class.

  For numeric metadata, z-score normalization was applied. Standardization for a dataset involves obtaining values with a mean of zero and a standard deviation of one. Therefore, if the mean of the original data is μ and its standard deviation is σ, the value z is calculated based on the following formula:

$$Z = \frac{x - \mu}{\sigma} \tag{3-1}$$

- **Removal of Unconventional Unicode**
  After normalizing the data, unconventional Unicode characters, such as symbols for country flags, are removed from the text.
- **Removal of Duplicate Samples or Those Without User Information**
  In the final preprocessing step, samples that are duplicated or lack metadata and user information are removed from the dataset.

After preprocessing steps, and the dataset will contain 7,545 samples.

In this section, we present our model named MDRD: A Deep **M**ulti-domain **R**umors **D**etection Model using the Mixture-of-Experts approach. The model architecture is shown in the figure 2.

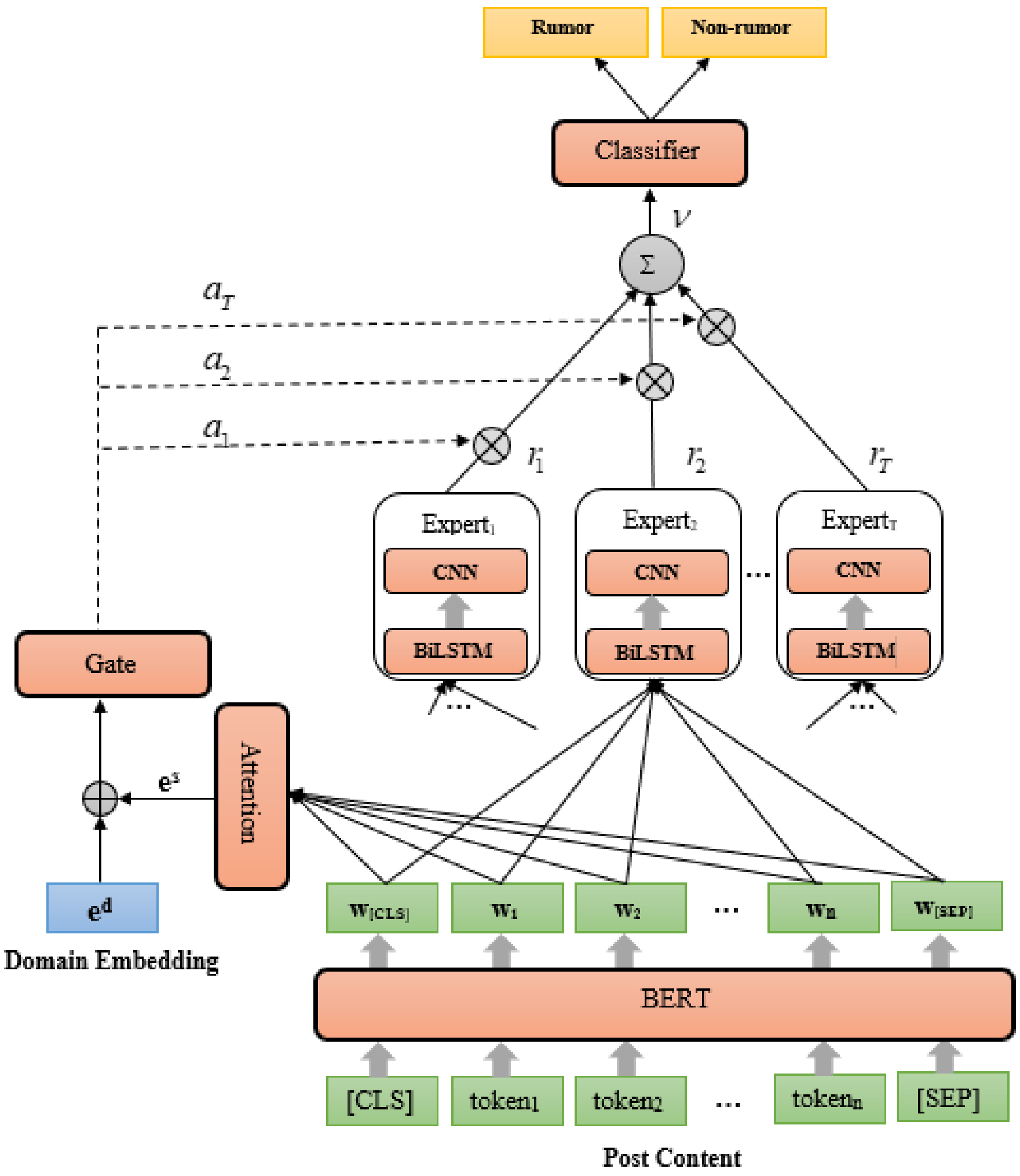

*Figure 2- The proposed MDRD framework for multi-domain rumor detection*

## 4-1- Word Embedding and Domain Embedding Extraction

In this study, the FarSSiBERT language model is used to obtain word embeddings. After preprocessing the dataset, the content is tokenized into constituent units using BertTokenizer, which employs the WordPiece algorithm. This algorithm includes specific tokens from the BERT model, such as [PAD], [UNK], [CLS], [MASK], and [SEP]. Additionally, suffixes are marked with the [##] symbol and separated from the root word as individual tokens. After adding special classification ([cls]) and separation ([sep]) tokens to the text, we obtain a list of units $[[CLS], token_1, ..., token_n, [SEP]]$ where *n* represents the number of tokens in the textual content. These tokens are fed into BERT to obtain word embeddings $W = [w_{[CLS]}, w_1, ..., w_n, w_{[SEP]}]$. To extract more effective features for each word, the average of the last four hidden layers' outputs is used as the text feature vector instead of directly using the [CLS] classification vector. These word embeddings are then processed through an attention-based network to obtain sentence-level embeddings $e^s$. To handle domain-specific customization, a learnable vector, $e^d$, known as the domain embedding, is defined to personalize the representation for each domain. Consequently, a unique domain-specific value $e^d$ is learned for each domain.

## 4-2- Representation Extraction

Due to the advantages of the Mixture-of-Experts method, we use multiple experts (networks) to extract various rumor representations. While a single expert can be used to extract rumor representations across multiple domains, each expert specializes in a specific domain and performs well in extracting features unique to that domain. Therefore, a representation extracted by a single expert contains only partial domain-specific information and cannot fully capture the features of rumor content[36]. As a result, this study employs multiple experts to ensure comprehensive representation.

An expert network can be denoted by $\Psi_i(W;\theta_i)(1 \le i \le T)$ where $W$ represents a set of word embeddings used as input to the expert network, $\theta_i$ denotes learnable parameters, and $T$ is a hyperparameter that specifies the number of expert networks. The output $r_i$ of an expert network represents a feature extraction output, such as a learned representation. Thus, we have the following equation:

$$r_i = \Psi_i(W;\theta_i), \qquad (4\text{-}1)$$

In this study, each expert network consists of a combination of two BiLSTM layers and a TextCNN layer, with additional metadata concatenated. Word embeddings are first passed as input to the LSTM, and the LSTM output is used as input for the CNN. The BiLSTM layer captures sequential and contextual dependencies from the input text and better handles long-range dependencies. The contextual output from BiLSTM is then passed to the CNN layer, which excels at identifying local patterns and features in the text.

Local features refer to patterns or structures within a limited portion of the text, such as specific word combinations or phrases indicative of a particular class. CNN is superior at capturing these local patterns due to its convolutional operations, enabling it to identify features like distinctive word pairs or phrases. Conversely, global features represent broader contextual understanding or overall structure, such as the meaning of a sentence influenced by paragraph or document level context. BiLSTM is well-suited for capturing long-range dependencies since it processes text sequentially and incorporates both preceding and succeeding word information, providing a better grasp of global context. Therefore, combining these two deep learning models enhances the proposed model's ability to understand both global and local aspects of the text.

In addition to textual content features, metadata such as post-related information and user details can play a critical role in rumor detection. Consequently, this study incorporates metadata, including the number of repost, the user's follower count, and the user's account creation date on X, by concatenating these features to the TextCNN output.

## 4-3- Domain Gate

To achieve good performance in multi-domain rumor detection, it is essential to generate high-quality rumor representations that can effectively capture rumors from different domains. A simple solution is to compute the average of the representations obtained by all experts. However, simple averaging results in the loss of domain-specific information, meaning that the generated representation may not perform well for multi-domain rumor detection. It is important to note that different experts specialize in different domains. To detect rumors, we adaptively select the experts. To this end, a **domain gate** that takes domain and sentence embeddings as input is proposed to guide the selection process. The output of the selection process is a vector $a$ that indicates the weight of each expert[36]. The domain gate is represented as $G(.;\phi)$ where $\phi$ is the parameters in the domain gate:

$$a = soft\max(G(e^d \oplus e^s;\phi)), \qquad (4\text{-}2)$$

Here, the domain gate $G(.;\phi)$ is a feed-forward network, and $e^d$ and $e^s$ represent the domain embedding and sentence embedding, respectively. The **softmax** function is applied to normalize the output $G(.)$, and weight vector $a \in \mathbb{R}^n$ that indicates the importance of each expert in detecting the rumor sample. With the domain gate, the final feature vector of the post will be:

$$\upsilon = \sum_{i=1}^{T} a_i r_i, \qquad (4\text{-}3)$$

### 4-4- Prediction

The final feature vector of the posts for rumor detection is fed into the classifier, which is a multi-layer perceptron network with a softmax output layer:

$$\hat{y} = soft\max(MLP(\upsilon)), \qquad (4\text{-}4)$$

The goal of the rumor detection model is to determine whether a post is a rumor or not. Here, $y^i$ is used to represent the true label and $\hat{y}^i$ to represent the predicted label. The binary cross-entropy loss (BCELoss) function is used for classification:

$$L = -\sum_{i=1}^{N}(y^i \log \hat{y}^i + (1-y^i)\log(1-\hat{y}^i)). \qquad (4\text{-}5)$$

This loss function is used for binary classification tasks and measures the difference between the predicted and true class labels. During model training, minimizing this difference leads to better accuracy in the final result. Figure 3 illustrates the MDRD model's core algorithm.

**Algorithm 1** Training Procedure of the MDRD

```
Require: Dataset D = {(p_j, y_j, d_j, m_j)}_{j=1}^N where p_j is a post, y_j is the label, d_j is the domain, m_j is a
    metadata vector; number of experts T; learning rate η;
Ensure: Trained MDRD model parameters Θ
  Initialize T BiLSTM–CNN experts {Ψ_i}_{i=1}^T with parameters {θ_i}_{i=1}^T, Domain Gate G(·; φ) and
    classifier MLP(·; ω)
  Θ ← {θ_1, ..., θ_T, φ, ω}
  for each epoch do
      for each mini–batch B = {(p_j, y_j, d_j, m_j)}_{j=1}^B do
          W_j ← BERT(p_j)
  // Expert encoding with metadata
          for i ← 1 to T do
              h_ij ← BiLSTM_i(W_j)
              c_ij ← CNN_i(h_ij)
              r_ij ← [c_ij || m_j]                                   ▷ concat metadata
          end for
  // Domain–gated fusion
          e_{d_j} ← DomainEmbed(d_j)
          e_{s_j} ← MaskAttention(W_j)
          a_j ← softmax(G([e_{d_j} || e_{s_j}]; φ))
          v_j ← Σ_{i=1}^T a_ij r_ij
  // Prediction
          ŷ_j ← softmax(MLP(v_j; ω))
      end for
      L ← (1/B) Σ_{j=1}^B BCE(ŷ_j, y_j)
      Update Θ ← Θ − η∇_Θ L                                          ▷ Adam
  end for
  return Θ
```

*Figure 3- Pseudo-code showing the algorithm for training MDRD model.*

# 5- Experment

## 5-1- Baseline Methods

This section reports a comparative evaluation between the proposed model and a set of state-of-the-art baselines. For the sake of fairness, each method was executed with its recommended parameters settings in the paper.

- The BERT_all[45] method utilizes the FarSSiBERT[12] language model together with a single-layer MLP classifier, trained simultaneously on a mixture of all domains—similar to training on a single, general-purpose dataset. We freeze all BERT layers and train the model for 50 epochs
- The EANN[37] model is designed for multimodal fake news detection and consists of three main modules: a feature extraction unit, an event discriminator, and a fake news classifier. For the purpose of our experiments, we focus exclusively on the textual feature extraction component. To use the model for Persian language, we use FastText embeddings with dimension set to 300.
- The MDFEND[36] is a multi-domain fake news detection model that uses a Mixture-of-Experts method to extract news features. It combines the outputs of multiple expert networks to produce the final news representations.

## 5-2- Experiment Setting

We divide our dataset into training, validation, and test sets using a 6:2:2 ratio. For a fair comparison, final results are averaged over ten independent runs. All experiments are conducted on an A100 GPU with 40 GB of memory. We use Python 3.8 and the PyTorch framework to implement our model. The maximum sentence length is set to 170, and BERT embeddings are 768-dimensional. The model is optimized using the Adam optimizer with a learning rate of 5e-4 and a batch size of 64. Detailed configurations are provided in Table 4. We report accuracy (ACC) and F1-score as evaluation metrics.

Table 4- The details of the parameters

| Hyperparameter | Value |
|---|---|
| Number of experts | 7 |
| Learning rate | 5e-4 |
| Optimizer | Adam |
| Weight decay | 5e-5 |
| Maximum sequence length | 170 |
| Word embedding dimension | 768 |
| Batch size | 64 |
| Maximum number of epochs | 50 |
| MLP dropout | 0.4 |

## 5-3- Results

In this section, we present and analyze the experimental results of our proposed model on the Persian dataset. Table 5 compares the performance of our proposed model against several baseline models, where the best results in each column are highlighted in bold.

Table 5- Experimental results on PerFact dataset

| Method | F1 | ACC |
|---|---|---|
| BERT_all | 70.60 | 70.77 |
| EANN | 71.01 | 71.30 |
| MDFEND | 78.07 | 78.18 |
| MDRD | **79.86** | **79.98** |

Table 6 presents the accuracy and F1-score obtained from the evaluation in each of the domains separately.

Table 6- Experimental results on the PerFact dataset, divided by domain

| Domain | F1 | ACC |
|---|---|---|
| Politics | 77.58 | 77.63 |
| Technology | 74.30 | 75.56 |
| Economy | 72.28 | 78.65 |
| Society | 85.58 | 85.88 |
| Health | 79.81 | 79.98 |
| Entertainment | 74.03 | 84.16 |
| Military | 61.57 | 90.10 |

Since the performance of most rumor detection methods drops significantly when faced with new events and unseen datasets, the proposed method in this study was also evaluated in a scenario involving a new event. In this context, the event related to the "Babak Khorramdin murder" from the social domain, with 449 samples, was completely excluded from the training and validation sets and used solely as the test set. Table 7 shows the results of comparing the proposed model and the baseline model in the described scenario, indicating that the proposed method performs better.

*Table 7- Experimental results on the PerFact dataset, with the "Babak Khorramdin" event used as test and the remaining data for training and validation"*

| **Method** | **F1** | **ACC** |
|---|---|---|
| MDFEND | 64.10 | 69.61 |
| MDRD | **66.84** | **69.63** |

### 5-4- Performance Analysis

The proposed model achieved an F1 score of 79.86% and an accuracy of 79.98%, outperforming baseline models. It effectively leverages data-rich domains to enhance rumor detection in data-scarce domains, reducing the need for retraining when new domains are introduced. The architecture—combining BiLSTM and TextCNN—improves both contextual and local feature extraction, while integrating user and post information enables the model to utilize both linguistic and social signals. This approach not only boosts accuracy but also strengthens the model's generalization and robustness to new, unseen rumors and events.

As shown in Table 6, the F1 score is lower in the military domain due to a strong imbalance—153 rumor samples vs. only 14 non-rumor samples. This highlights the importance of balanced data within each domain to avoid overfitting and ensure fair learning, which can be addressed by techniques like data augmentation or targeted data collection.

## 5-5- Hyper Parameters Sensivity

In this section, we examine the sensitivity of several hyperparameters, including the MLP classifier's dropout rate, maximum sequence length, and learning rate. Figure 3 shows that our model achieves its best performance with a dropout of 0.4, max sequence length of 170, and learning rate of 5e−4, reaching an F1 score of 0.7986.

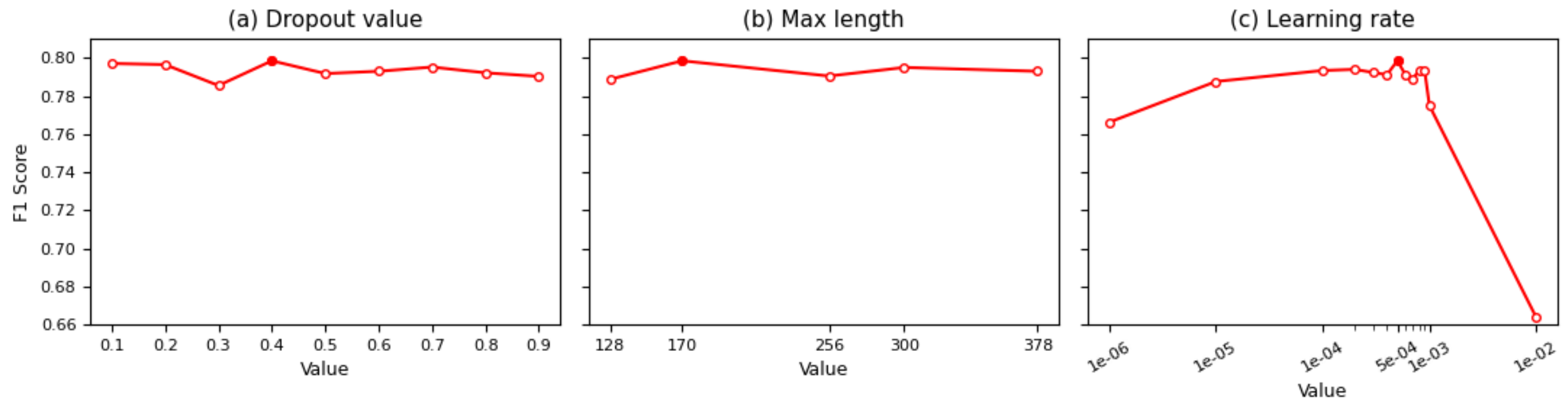

*Figure 4- Performance of MDRD model with various hyperparameters*

## 5-6- Ablation Study

In this section, we analyze the effectiveness of different components of our proposed model through ablation experiments by removing or changing specific variables. We define the MDRD variants as follows:

• w/o LSTM: This variant excludes the BiLSTM layer from the expert architecture.
• w/o Metadata: In this setup, no metadata (such as repost count, user follower count, or account creation date) is provided to the experts.
• We also experiment with different BERT embedding strategies for posts, including using only the last hidden layer, the mean of the last two layers, and the mean of the last three layers, instead of our recommended setting (the mean of the last four layers).
• Additionally, we replace FarsiBERT with other multilingual and monolingual models that support Persian, namely LaBSE and ParsBERT v2.

Table 8 shows the performance of these variants on the PerFact dataset.

*Table 8- Experimental results on PerFact dataset, for each different variant*

| Method | F1 | ACC |
|---|---|---|
| w/o LSTM | 0.7918 | 0.7930 |
| w/o Metadata | 0.7920 | 0.7935 |
| Last hidden layer | 0.7877 | 0.7890 |
| Mean of  last two layers | 0.7952 | 0.7967 |
| Mean of last three layers | 0.7944 | 07954 |
| LaBSe | 0.7743 | 0.7752 |
| ParsBERT | 0.7773 | 0.7782 |
| MDRD | **0.7986** | **0.7998** |

The results show that our proposed model outperforms the other variants, highlighting not only the benefits of adding the LSTM layer but also the value of incorporating metadata, which provides additional information crucial for rumor detection. We also see a clear advantage when using the optimal setup for capturing textual information.

# 6 – Discussion

This study provides clear answers to its research questions. First, the proposed hybrid model demonstrates strong performance in detecting multi-domain rumors, achieving an F1 score of 79.86%, which reflects its robust accuracy across diverse topics. Second, the newly collected dataset from the X platform, annotated with multiple rumor veracity labels spanning seven thematic domains, has been validated with a Fleiss' Kappa of 0.74, indicating satisfactory reliability for training and evaluation purposes.

Although the dataset used in this work is Persian, the architecture and methodological framework are language-independent and can be adapted to other social media contexts and languages with minimal modification.

Despite these promising outcomes, the research faced notable challenges, primarily related to dataset collection. Due to restrictive policies imposed by the X platform's API, data access for recent events was limited, which constrained the volume and diversity of the dataset. Additionally, the limited number of active Persian fact-checking websites—critical for identifying and verifying rumor events—further complicated data acquisition. Consequently, finding events with comprehensive fact-checking and sufficient data proved difficult, limiting the dataset's potential coverage.

These challenges underscore key areas for improvement. To enhance the overall performance of the proposed model, several promising research directions are recommended. Since textual content often spans multiple domains, incorporating multi-label domain annotations for each sample, acknowledging that individual samples may belong to multiple overlapping domains, could enrich the model's contextual understanding and improve its generalizability across topics. Furthermore, exploring alternative text feature extraction methods, such as TextGNN, which may result in richer semantic embeddings, offers a promising avenue beyond the current hybrid of TextCNN and BiLSTM, potentially advancing rumor detection capabilities.

# 7-Conclusion

This study addresses the challenge of detecting multi-domain rumors on social media by proposing a novel hybrid approach based on a Mixture-of-Experts framework enhanced with a domain gating mechanism. Leveraging transformer-based FarSSiBERT embeddings, each expert combines deep learning architectures to effectively capture semantic and contextual features across diverse topics. Experimental results demonstrate that integrating convolutional and recurrent neural networks within experts, alongside metadata utilization, significantly improves classification accuracy and robustness in rumor detection. The proposed model achieves an F1 score of 79.86% and accuracy of 79.98%, indicating strong performance compared to existing methods. Furthermore, this research introduces a new annotated dataset of Persian social media posts, encompassing multiple rumor veracity classes across seven thematic domains, validated with an acceptable inter-annotator agreement (Fleiss' Kappa = 0.74). Overall, this work contributes both an effective multi-domain rumor detection methodology and a valuable dataset, advancing the study of misinformation in digital discourse.

## Declaration of Computing Interest

The authors declare that they have no known competing financial interests or personal relationships that could have appeared to influence the work reported in this paper.

## Data availability

The source code and data for this research will be made publicly available on GitHub[5] upon the acceptance of this paper.

## Author Contributions Statement

The conception and design of the study, along with material preparation, data collection, and analysis, were carried out collaboratively by all authors. The first draft of the manuscript was written by Mohadeseh Sheikhqoraei. All authors discussed the results and contributed to the final manuscript.

## Corresponding author

Correspondence to Zainabolhoda Heshmati.

[5] https://github.com/Mqoraei